\documentclass[sigconf, nonacm, natbib]{acmart}

\usepackage{adjustbox}
\usepackage{amsmath,bm,bbm}
\usepackage[ruled,lined,linesnumbered, commentsnumbered, longend]{algorithm2e}
\usepackage[justification=justified,skip=3pt]{caption}
\usepackage{pgfplots}
\usepackage{multirow}
\usepackage{tikz}
\usetikzlibrary{matrix}
\usepackage{amsmath}

\makeatletter
\def\BState{\State\hskip-\ALG@thistlm}
\makeatother
\usepackage{tkz-euclide}
\usetikzlibrary{shapes.arrows, fadings, automata,tikzmark,decorations.pathreplacing,patterns}
\usetikzlibrary{intersections}
\usetikzlibrary{arrows.meta}
\usetikzlibrary{positioning, fit, mindmap, trees, calc,tikzmark,shapes}
\begin{document}

\title{Context-Aware Lifelong Sequential Modeling for Online Click-Through Rate Prediction}


\author{Ting Guo}
\affiliation{%
  \institution{Wechat Channels, Tencent}
  \city{Shenzhen}
  \country{China}
}
\email{tinnaguo@tencent.com}

\author{Zhaoyang Yang}
\authornote{corresponding author}
\affiliation{%
  \institution{Wechat Channels, Tencent}
  \city{Shenzhen}
  \country{China}
}
\email{terrellyang@tencent.com}

\author{Qinsong Zeng}
\affiliation{%
  \institution{Wechat Channels, Tencent}
  \city{Guangzhou}
  \country{China}
}
\email{qinzzeng@tencent.com}

\author{Ming Chen}
\affiliation{%
  \institution{Wechat Channels, Tencent}
  \city{Guangzhou}
  \country{China}
}
\email{mingchen@tencent.com}

\renewcommand{\shortauthors}{Ting Guo et al.}
\renewcommand{\shortauthors}{}

\newcommand{\reed}[1]{{\color{red}{(reed: #1)}}}
\newcommand{\tbd}[1]{{\color{red}{(#1)}}}
\begin{abstract}

Lifelong sequential modeling (LSM) is becoming increasingly critical in social media recommendation systems for predicting the click-through rate (CTR) of items presented to users. Central to this process is the attention mechanism, which extracts interest representations with respect to candidate items from the user sequence. Typically, attention mechanisms operate in a point-wise manner, focusing solely on the relevance of individual items in the sequence to the candidate item. In contrast, context-aware LSM aims to also consider adjacent items in the user behavior sequence to better assess the importance of each item. In this paper, we propose the Context-Aware Interest Network (CAIN), which utilizes the Temporal Convolutional Network (TCN) to create context-aware representations for each item throughout the lifelong sequence. These enhanced representations are then used in the attention mechanism instead of the original item representations to derive context-aware interest representations. Building upon this TCN framework, we propose the Multi-Scope Interest Aggregator (MSIA) module, which incorporates multiple TCN layers and their corresponding attention modules to capture interest representations across varying context scopes. Furthermore, we introduce the Personalized Extractor Generation (PEG) module, which generates convolution filters based on users' basic profile features. These personalized filters are then used in the TCN layers instead of the original global filters to generate more user-specific representations. We conducted extensive experiments on both a public dataset and an industrial dataset from the WeChat Channels platform. The results demonstrate that CAIN outperforms existing methods in terms of prediction accuracy and online performance metrics. 

\end{abstract}


\maketitle

\section{introduction}

Click-through rate (CTR) prediction is a fundamental task for recommendation systems on today's social media platforms. The objective is to predict the likelihood that a user will click on an item presented to them. The accuracy of this prediction heavily depends on understanding the user's interests with respect to the candidate items.

In recent years, deep neural networks (DNNs) have significantly improved CTR prediction accuracy in most scenarios. The key to this success lies in the ability of these networks to model users' historical behavior sequences. Central to this process is the attention mechanism, which provides a relevance score between each item in the sequence and the candidate items. These scores, known as attention scores, are then used to perform a weighted summation of the sequence to generate the final interest representations for the candidate item. Much work has been done to improve the efficiency of attention mechanisms \cite{zhou2018deep, zhou2019deep}.

As user behavior becomes richer, the length of the user's historical behavior sequence grows dramatically, potentially extending to a lifelong span. This increase in sequence length leads to a significant rise in the computational cost of the attention. An efficient approach to mitigate this computational burden in lifelong sequential modeling (LSM) is to divide the attention mechanism into two units: a General Search Unit (GSU) and an Exact Search Unit (ESU) \cite{pi2020search}. The GSU's role is to sift through the lifelong sequence to identify items that are most relevant to the candidate items. Subsequently, the ESU extracts user interest representations from the items identified by the GSU. This division allows the model to handle much longer sequences, which contain richer information about user interests, thereby further improving CTR prediction accuracy. 

\begin{figure}
    \centering
    \includegraphics[width=0.95\linewidth]{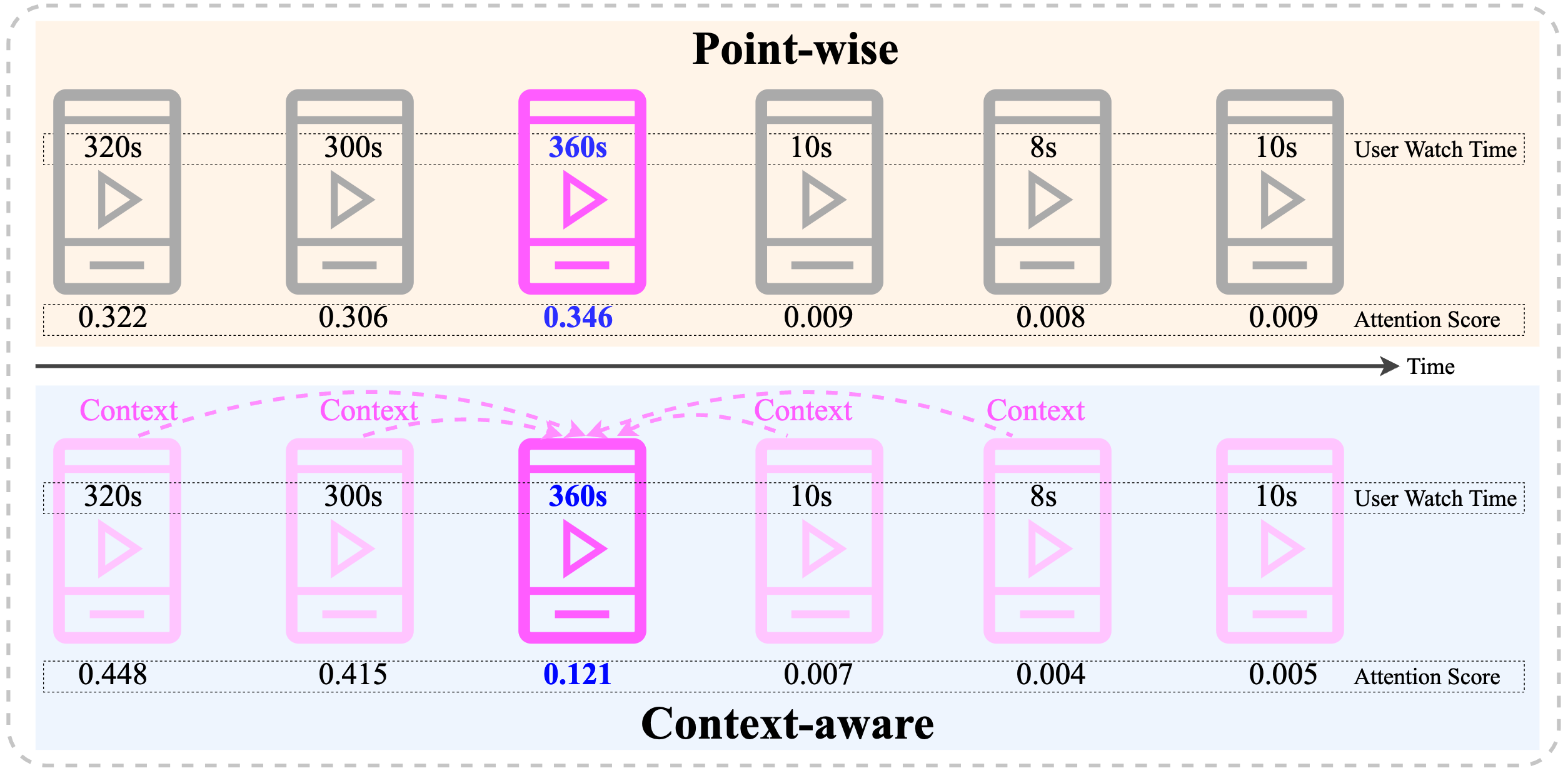}
    \caption{A comparison of attention scores from point-wise and context-aware perspectives. }
    \label{fig:contextAware}
    \vspace{-0.5cm}
\end{figure}

However, most previous work considers attention as a point-wise scoring process, typically analyzing the relevance between candidate items and each individual item in the sequence. This approach overlooks the valuable information that adjacent items can provide in understanding user intent. This is particularly important on today's social media platforms such as TikTok, YouTube, and WeChat Channels, where users often consume a series of items consecutively. For instance, as illustrated in Figure \ref{fig:contextAware}, in a sequence of user behaviors generated from a consecutive consumption flow, the user has a very long watch time for the first three items, but the watch time starts to decrease dramatically from the fourth item onward. From a point-wise perspective, the third item might appear to be satisfactory since the user watched it for a long duration. However, from a context-aware perspective, the third item may not be as favorable because it is the last item the user was willing to watch extensively. There may be certain properties of the third item that caused the user to lose interest, leading to a decrease in watch time for subsequent items. Therefore, it is crucial for the model to employ context-aware LSM, which considers the context information contained in the adjacent items of each item in the sequence. This approach ensures a more comprehensive understanding of user interests with respect to candidate items and can provide more continuous and relevant recommendation outcomes.

To achieve context-aware LSM, we propose the Context-Aware Interest Network (CAIN). CAIN first extracts context-aware representations of each item in the historical behavior sequence by performing a convolution operation along the temporal axis of the sequence, known as the Temporal Convolutional Network (TCN) \cite{bai2018empirical}. To the best of our knowledge, this is the first network to incorporate TCN in LSM. TCN offers two major advantages: it is lighter and more computationally efficient compared to methods such as Recurrent Neural Networks (RNN) \cite{hochreiter1997long,chung2014empirical} and Self-Attention \cite{vaswani2017attention}, and the context length can be easily controlled by adjusting the filter size of the convolution. The output representations from the TCN are then used in the subsequent attention module, instead of the original item representations, to extract context-aware interest representations with respect to the candidate items. Building upon this TCN framework, we have embedded a Multi-Scope Interest Aggregator (MSIA) module in CAIN. The MSIA module contains multiple stacked TCN layers that gradually extend the receptive field of the output representations. The output of each layer is sent to its corresponding attention modules to extract interest representations of different context scopes with respect to the candidate items. The computational cost of the attention decreases for the latter convolution layers as the sequence length is reduced through the TCN layers. Finally, to enhance the personalization capability of the convolution operations, we propose the Personalized Extractor Generation (PEG) module. This module generates convolution filters for different users based on their basic profile features. Instead of using uniform convolution filters for all users, we use the filters generated by the PEG module in all TCN layers. This makes the output representations of the TCN more user-specific, which can further improve the representativeness of the final interest representations.

We conducted extensive experiments on both a public dataset and an industrial dataset collected from user traffic logs on the WeChat Channels platform. The results indicate that the proposed CAIN achieves higher CTR prediction accuracy compared to existing methods. Additionally, the results show that the TCN framework and the MSIA module are highly adaptive and provide performance gains over multiple LSM baselines with different attention designs. Notably, CAIN also achieved significant improvements in online A/B tests. These findings demonstrate the effectiveness and robustness of CAIN in enhancing CTR prediction in complex environments.

\section{related work}

\subsection{Sequential Modeling}

Deep learning based models have achieved significant improvements in industrial applications, such as online advertising and recommendation systems \cite{cheng2016wide, qu2016product, wang2017deep, guo2017deepfm, wang2021dcn, lian2018xdeepfm, song2019autoint, zhou2018deep, pi2020search, xiao2017attentional}. The modeling of the user's historical behavior sequence, known as sequential modeling (SM), is crucial for these models to understand user intent and achieve personalized predictions. Much work has been done in this area \cite{zhou2018deep, pi2020search, zhou2019deep, feng2019deep, chen2022efficient, chang2023twin}.

As user behavior becomes more complex, the length of historical behavior sequences grows dramatically. Consequently, more work has focused on lifelong sequential modeling (LSM) in recent years. Sim \cite{pi2020search} and UBR4CTR \cite{qin2020user} are two methods that introduce a two-stage framework to model user lifelong sequences, consisting of a General Search Unit (GSU) and an Exact Search Unit (ESU). The GSU retrieves the top-k items most relevant to the target item from the entire user behavior history, which are then fed into the ESU for subsequent attention. This framework heavily relies on pre-trained embeddings, which can reduce the consistency between the GSU and ESU stages. To address this, ETA \cite{chen2022efficient} was proposed to use SimHash \cite{charikar2002similarity} to retrieve relevant items and encode item embeddings via locality-sensitive hash (LSH) in the ESU. SDIM \cite{cao2022sampling} was also proposed to generate hash signatures for both the candidate and behavior items, then gather behavior items with matching hash signatures to represent user interest. Both methods allow the two stages to share identical embeddings to increase consistency. Furthermore, TWINS \cite{chang2023twin} was proposed to enhance consistency by introducing CP-GSU, which retrieves behaviors that are not only target-relevant but also considered important by the ESU. Additionally, some work has upgraded the two-stage framework into a three-level attention pyramid \cite{hou2024cross} to further enhance consistency among stages.

However, most of this work considers LSM as a point-wise process, focusing solely on the relationship between individual items in the sequence and the candidate item. They overlook the importance of the context information provided by adjacent items in the sequence. In this paper, we aim to achieve a context-aware LSM that takes into account the adjacent items of an item in the sequence.

\subsection{Context-Aware Modeling}

Context-aware modeling methods are widely used in the fields of natural language processing (NLP), computer vision (CV) and speech recognition (SR). Long Short-Term Memory (LSTM) networks \cite{hochreiter1997long} and Gated Recurrent Units (GRUs) \cite{chung2014empirical} are classic RNN models that are extensively utilized in various NLP and SR tasks \cite{yin2017comparative,8690387,liu2016recurrent,graves2013speech,graves2014,mirsamadi2017automatic}. In recent years, the Transformer model and Self-Attention mechanisms \cite{vaswani2017attention} have become fundamental components in NLP, featuring an encoder and decoder based solely on attention mechanisms. GPT \cite{radford2018improving} and BERT \cite{devlin2018bert} are two well-known models built on this module. Besides RNN-based and attention-based approaches, other methods for context-aware modeling also exist. The Temporal Convolutional Network (TCN) \cite{bai2018empirical} is one such option. Convolutional Neural Networks (CNNs) dominate CV tasks \cite{krizhevsky2012imagenet, simonyan2014very, girshick2014rich, girshick2015fast, ren2015faster, redmon2016you, he2016deep}, as the convolution operation is well-suited for processing images to generate high-level feature maps by inherently considering the neighboring pixels of the target. TCNs conduct the convolution operation along the temporal dimension, enabling the generated representations to be context-aware \cite{Lea_2017_CVPR,Lea_2016_ECCV,Farha_2019_CVPR,8683634,Cheng_2020_ECCV}.

Researchers have incorporated RNNs and Self-Attention in SM for CTR prediction. CA-RNN \cite{liu2016context} and CRNNs \cite{smirnova2017contextual} are two methods that use RNNs to predict the probability of the next item given the user's historical items. DEIN \cite{zhou2019deep} employs GRUs to extract each user's interest state and utilizes AUGGRU to model the interest evolution with respect to the target item. Considering that sequences are composed of sessions, DSIN \cite{feng2019deep} uses a Self-Attention mechanism to extract users’ interests in each session and then applies Bi-LSTM to model how users’ interests evolve across sessions.

However, although these methods have brought certain improvements to the CTR task, they face challenges in scaling to LSM due to the heavy computational burden. Therefore, in the LSM setting, a lighter and more computationally efficient approach for capturing context information is required.

\section{preliminaries}

In this paper, we explore the topic of context-aware lifelong sequential modeling (LSM). Unlike traditional LSM, which typically treats items in a sequence as isolated behaviors during the attention process to assess their relevance to candidate items, a context-aware LSM approach considers the context of each item within the sequence. The context, as referred to in this paper, consists of the adjacent items surrounding each item in the sequence, typically representing the items the user interacted with immediately before and after it.

To formalize, let $\vec{LH} = \{ lh_1, lh_2, \cdots, lh_{N} \}$ represents the lifelong behavior sequence of a user. The context of the item $lh_{t}$, referred to as the center item, can be denoted as:
\begin{equation} 
    \vec{cxt_t} = \{ lh_{t-cl}, lh_{t-cl+1}, \cdots, lh_{t-1}, lh_{t+1}, \cdots, lh_{t+cl-1}, lh_{t+cl} \},
\label{eq:context}
\end{equation}
where $cl$ represents the context length.

For all models presented in this paper, we mainly include three distinct categories of features for each user:

\begin{itemize}
 \item Basic profile features, denoted as ${\{B\}}$. 
 \item The short-term behavior sequence of the user, represented by $\vec{H} = \{ h_1, h_2, \cdots, h_{N} \}$. 
 \item The lifelong behavior sequence of the user, represented by $\vec{LH} = \{ lh_1, lh_2, \cdots, lh_{N} \}$. 
\end{itemize}

Note that each item $lh_{N}$ or $h_N$ in the lifelong and short-term behavior sequences includes its item ID and side information, such as the user's watch time. For a given user-item pair <$u_{i}$, $v_{i}$>, the model aims to predict the click-through rate (CTR) for user $u_{i}$ with respect to item $v_{i}$ as follows:
\begin{equation} 
    p_{i} = P(y_i=1|~u_{i},v_{i},B,\vec{H}^{t},\vec{LH}^{t};\theta),
\label{eq:target}
\end{equation}
where $\theta$ represents the parameters of the model.

The main network is optimized using a cross-entropy loss function, which is defined as follows:
\begin{equation}
\begin{split}
    \mathcal{L}_{CTR} = - \frac{1}{BS} \sum_{i=1}^{BS} & ~ \Big(~ y_i*log(p_i) + (1-y_i)*log(1-p_i) \Big),
\end{split}
\label{eq:cross_entropy}
\end{equation}
where $y_{i} \in \{0, 1\}$ represents the actual user feedback, and $BS$ denotes the total number of sample pairs in a training batch.

\section{methodology}

In this paper, we propose a novel Context-Aware Interest Network (CAIN) to enable the model to consider the context information of each item in the user's historical sequence when performing lifelong sequential modeling (LSM). The proposed CAIN incorporates the Temporal Convolutional Network (TCN) to calculate context-aware representations of the items in the sequence. These representations are then used in the subsequent attention module instead of the original item representations, allowing the model to extract context-aware interest representations with respect to the candidate items.

Based on this TCN framework, we further include two major modules in CAIN to enhance its performance: the Multi-Scope Interest Aggregator (MSIA) module and the Personalized Extractor Generation (PEG) module. The MSIA module consists of several TCN layers to extract representations within different context lengths, which are then fed into their corresponding attention modules to extract interest representations of varying context scopes. The PEG module contains a lightweight network to generate convolution filters based on the basic profile features of the user.

\begin{figure*}
    \centering
    \includegraphics[width=0.95\textwidth]{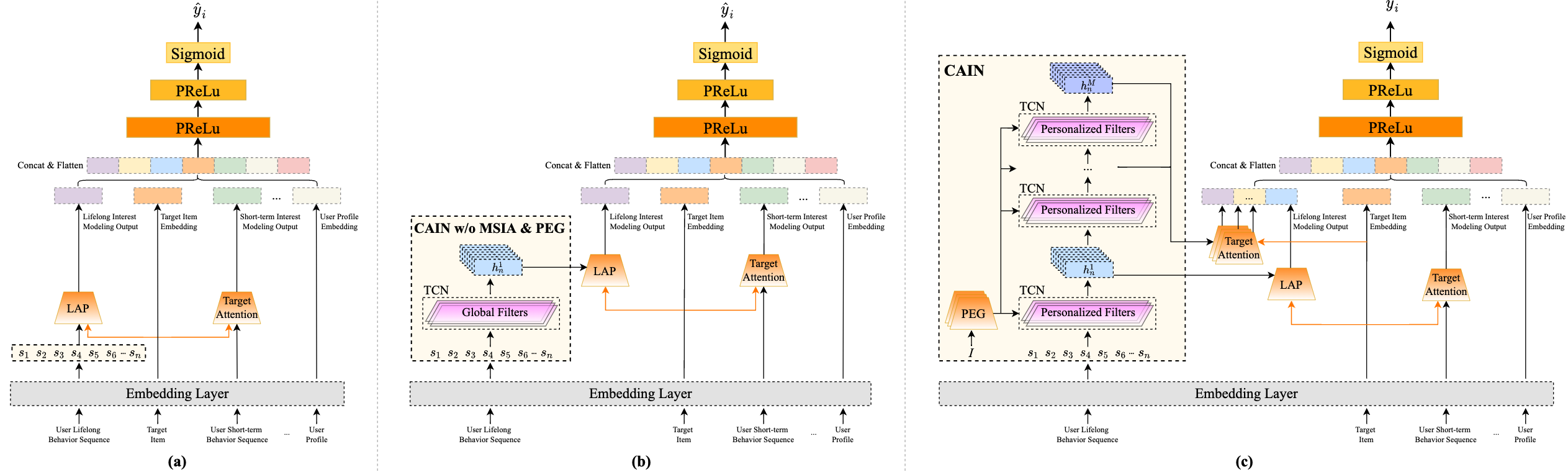}
    \caption{An overview of the proposed CAIN and its comparison to traditional LSM models. Figure (a) shows our baseline network that incorporates the LAP \cite{hou2024cross} for LSM. Figure (b) shows an upgraded model that incorporates the TCN framework proposed in CAIN. Figure (c) is the final version of the proposed CAIN, which includes the MSIA and PEG modules. }
    \label{fig:architecture}
    \vspace{-0.3cm}
\end{figure*}

We provide an overview of the proposed CAIN and its comparison to traditional LSM networks in Figure \ref{fig:architecture}. 

\subsection{Context Information Extraction}

A common approach to extract a user's interest with respect to candidate items from the lifelong sequence is to conduct an attention process. In this process, the candidate item, denoted as $v$, serves as the query to form pairs with the items in the sequence <$v, lh_t$>, where $lh_t \in \vec{LH}$. Each $<v, lh_t>$ pair is assigned an attention score based on the networks applied in the attention module:
\begin{equation} 
    s_t = Attn(ve, lhe_t;\theta_a),
\label{eq:attention}
\end{equation}
where $ve$ and $lhe_t$ represent the item representations of $v$ and $lh_t$, $theta_a$ represents the parameters in the attention module and $s_{i}$ is the attention score. These attention scores are then used in subsequent retrieval or weighted summation stages. 

It is worth noting that the above attention process primarily focuses on understanding the relationship between $v$ and $lh_t$. However, in most real-world scenarios, user behavior often occurs in a consecutive manner, where a user interacts with a series of items in succession. This underscores the importance of considering the context, specifically the items before and after each center item in the sequence, to fully understand the behavior and its relevance to the candidate items.

In the proposed CAIN, this context-aware LSM is achieved by utilizing the Temporal Convolutional Network (TCN) \cite{bai2018empirical} before the attention module. 

\subsubsection{Representation Extraction}

Compared to methods such as Recurrent Neural Networks (RNNs) \cite{hochreiter1997long, chung2014empirical} or Self-Attention \cite{vaswani2017attention}, TCNs offer two primary advantages. First, TCNs are lightweight and computationally efficient. They require only a single matrix multiplication to slide the convolution filters over the sequence. In contrast, RNNs and Self-Attention involve significantly more computation. In particular, RNNs include operations that cannot be parallelized. Second, TCNs allow for explicit control over the context length. In RNNs and Self-Attention, context length is typically learned implicitly during training, which can complicate generalization. We will explore this further in the Experiment section.

The TCN operation can be conceptualized as a linear layer moving through the sequence. Given a context length $cl$, the size of the convolution filter $W^C$ is $2*cl + 1$, encompassing the context on both sides of the center item $lh_{t}$. Let $ctxe_t$ represent the item representations within the context $ctx_t$, the calculation within each convolution window is expressed as:
\begin{equation} 
    cr_t = (ctxe_t \cup lhe_t) \times W^C + b_c,
\label{eq:conv}
\end{equation}
where $cr_t$ is the output context-aware representation of $lh_{t}$, and $b_c$ is the bias term. 

For the first and last few elements in the sequence that lack sufficient context items before or after them, we apply zero-padding to ensure they fit the specified context length. 

\subsubsection{Representation Substitution}

After the TCN, the original item representations $lhe_t$ are transformed into context-aware representations $cr_t$, which contain information about both the center item and its adjacent context items. To obtain a context-aware interest representation of the user with respect to the candidate item, we replace $lhe_t$ used in the attention module with $cr_t$. Consequently, equation \ref{eq:attention} is updated to:
\begin{equation} 
    s_t = Attn(ve, cr_t;\theta_a).
\label{eq:attention_c}
\end{equation}

Similarly, in other processes within the attention module, we use $cr_t$ instead of $lhe_t$. This modification ensures that the attention module considers the broader context of user interactions, leading to more accurate and context-aware interest representations. 

\subsection{Multi-Scope Interest Aggregator}

In many applications, different lengths of context provide varying insights into the sequence. For the context of items in the user's historical behavior sequence, a longer context length includes more profound influences of the item on the user, such as interest migration or degradation. In contrast, a shorter context length generally captures information about the user's immediate behavior changes after being presented with the item. Both types of information are crucial for fully understanding user intent. 

\begin{figure}
    \centering
    \includegraphics[width=0.95\linewidth]{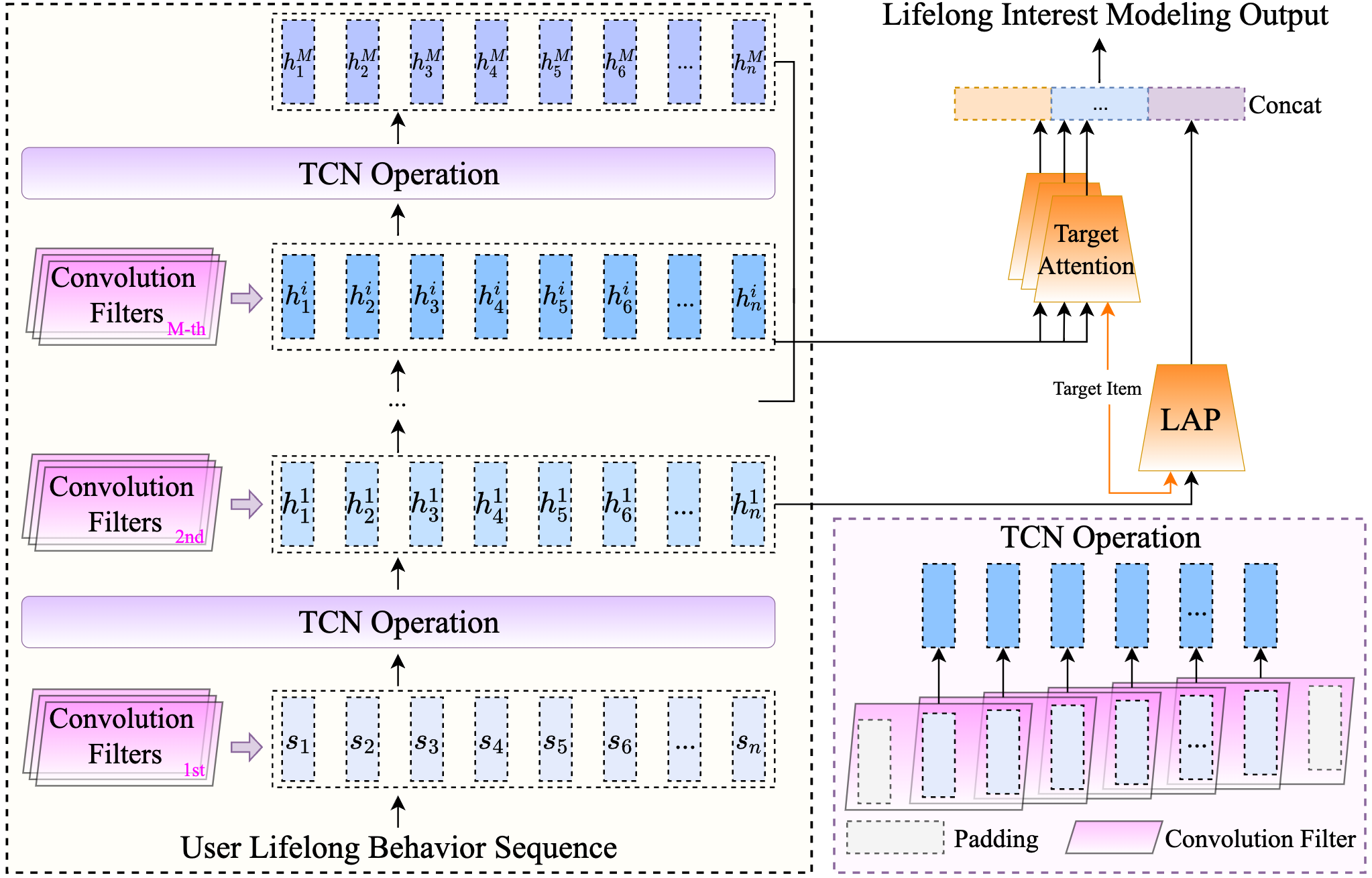}
    \caption{An illustration of the MSIA module. It contains several TCN layers and their corresponding attention modules to extract interest representations with respect to the candidate items within different context scopes. }
    \label{fig:MSIA}
    \vspace{-0.5cm}
\end{figure}

To extract context-aware interest representations under different context scopes, we propose the Multi-Scope Interest Aggregator (MSIA) module, which extends the above TCN framework. An illustration of the MSIA module is provided in Figure \ref{fig:MSIA}. 

\subsubsection{Stacking Layers}

The nature of TCN layers allows us to efficiently expand the context length by stacking more cascading layers. The receptive field of the output from the TCN layers gradually increases as the network deepens. Let $cl_n$ represent the context length of the $n-th$ layer, With a convolution stride size of 1, the context length of the $n+1-th$ layer is given by:
\begin{equation} 
    cl_{n+1} = cl_n + fs_{n+1} - 1,
\label{eq:receptive}
\end{equation}
where $fs_{n+1}$ is the filter size of the $n+1-th$ layer. 

As a result, the output of each layer contains information from different context lengths. This approach is more adaptive than tuning the context length of a single TCN layer, as it allows us to obtain various representations under multiple context scopes. At the same time, the representativeness of the output from the latter layers increases due to the non-linearity of the TCN operations, which also helps improve the final performance. 

To reduce the computational cost of processing the output from the deeper layers, the strides of the layers, except for the first layer, are set to values greater than 1. This significantly reduces the length of the output in these layers, thereby decreasing the computation required in their subsequent attention modules. Using a stride greater than 1 also accelerates the expansion of the context length, enabling the achievement of larger context scopes with fewer TCN layers. The stride of the first layer is fixed at 1 to ensure that each center item in the lifelong sequence has its own representation. With a stride greater than 1 in the subsequent layers, the information from multiple inputs is merged into a single output representation. Consequently, individual inputs no longer have their distinct representations in the output. This is acceptable for the deeper layers, as the information about individual items becomes less critical in longer context scopes. We will discuss this further in the Experiment section. 

\subsubsection{Individual Attentions}

To extract the interest representation of different context scopes with respect to the candidate items, we use attention modules with individual parameters to process the outputs of each TCN layer. This approach avoids inter-layer influence and allows for parallel computation. The attention modules employed can be categorized into two types: major attention and auxiliary attention. 

The major attention is applied only to the output of the first TCN layer to extract the interest representation $IR_1$ from the context-aware representations $cr^1_t$ of this layer. The attention technique should be suitable for a lifelong setting, as the length of the output of this layer matches the length of the input lifelong behavioral sequence. We apply the Lifelong Attention Pyramid (LAP) \cite{hou2024cross} for the major attention, as the LAP is used in our baseline. However, it is compatible with other lifelong attention mechanisms such as ETA \cite{chen2022efficient} and SDIM \cite{cao2022sampling}. 

The auxiliary attention is applied to the outputs of the subsequent TCN layers. Since the length of the output in these layers is much shorter than that of the first layer due to convolution strides greater than 1, and the focus is more on the information of items within the entire context scope rather than individual items, the attention mechanism used can be simpler than the major attention. We employ target attention with linear projection layers for auxiliary attention. Formally, the attention score $sa_t$ and the interest representation output $IR_n$ of the auxiliary attention for the $n-th$ layer are calculated using the following formulas: 
\begin{equation}
    sa_t = \frac{(W^Q_n \times ve) \times (W^K_n \times cr^n_t)\top}{\sqrt{d}},
\label{eq:attention_aux3}
\end{equation}
\begin{equation}
    IR_n = \sum_{t=1}^{T_n} sa^n_t*(W^V_n \times cr^n_t),
\label{eq:attention_aux4}
\end{equation}
where $W^Q_n$, $W^K_n$ and $W^V_n$ denote the projection weights, $d$ represents the inner dimension, and $cr^n_t$ is the representation output of the $n-th$ TCN layer. 

At the end of the MSIA module, all interest representations $IR_n$, including $IR_1$, are concatenated to form an integrated representation of multiple context scopes. This integrated representation captures a comprehensive view of user interests across different temporal scales, balancing the need for detailed individual item information and broader context understanding. 

\subsection{Personalized Extractor Generation}

Traditional convolution filters generally share parameters across different inputs, operating under the assumption that the inputs are drawn from very similar distributions. However, when using TCN to process users' historical sequences, this assumption may not hold. User behavior can vary significantly across different kinds of users, and the influence of an item on subsequent behavior can also differ greatly. For instance, a highly active user may decide to watch an item primarily based on its content, with minimal influence from previously presented items, even if some of those items were unsatisfactory. Conversely, a less active user, whose behavior is more susceptible to external factors, may be more influenced by the items presented before a given item. 

\begin{figure}
    \centering
    \includegraphics[width=0.9\linewidth]{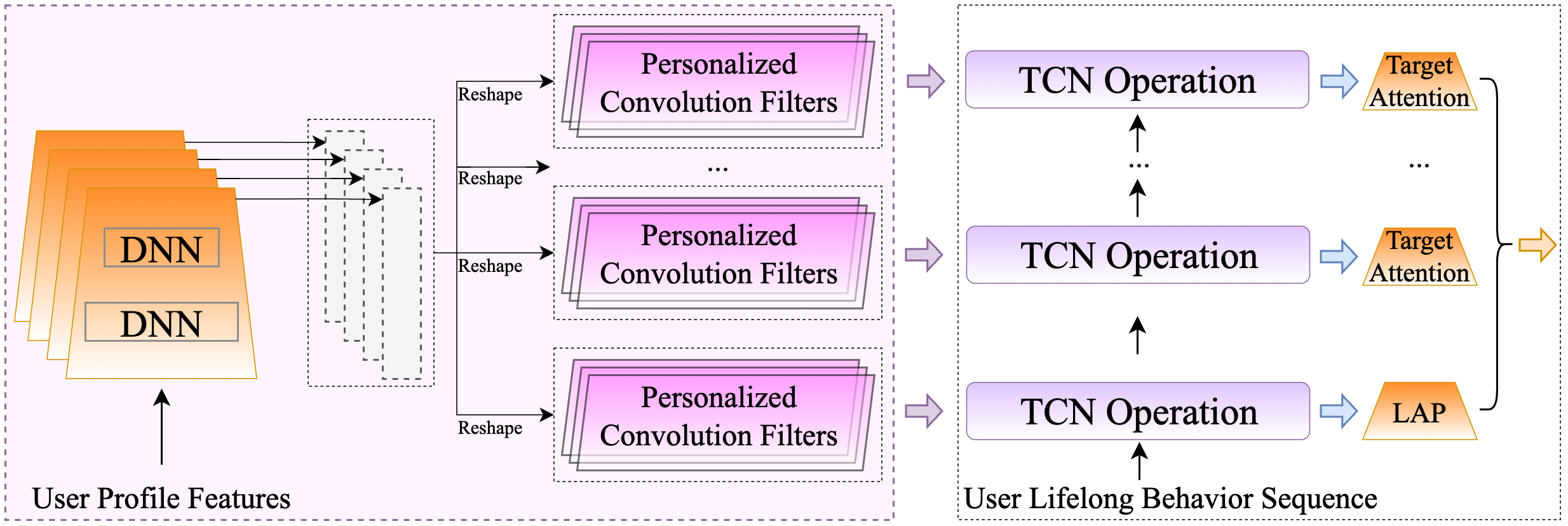}
    \caption{An illustration of the PEG module. It is a lightweight sub-network that takes the user's basic profile features as input and outputs the convolution filters that will be used in the TCN layers. }
    \label{fig:PEG}
    \vspace{-0.3cm}
\end{figure}

To make the context extraction more personalized, we propose a Personalized Extractor Generation (PEG) module, inspired by the work presented in the Personalized Cold Start Modules (POSO) \cite{dai2021poso}. The PEG module provides users with their own convolution filters, tailored to their unique behavior patterns. An illustration of the PEG module is presented in Figure \ref{fig:PEG}. 

\subsubsection{Filter Generation}

The PEG module contains a lightweight sub-network designed to generate personalized convolution filters $W^{PC}_n$ for each user. This sub-network consists of two fully connected layers that capture the interactions between the user's basic profile features and transform these inputs into the filter parameter space. The output of this sub-network is then reshaped to form the personalized convolution filters. Formally, the calculation of this generation process is as follows: 
\begin{equation}
    O_n=ReLU(W^{P1}_n \times I + b^{P1}_n),
\label{eq:peg1}
\end{equation}
\begin{equation}
    W^{PC}_n=Reshape(W^{P2}_n \times O_n + b^{P2}_n),
\label{eq:peg2}
\end{equation}
where $W^{P1}_n$ and $b^{P1}_n$ are the weights and biases of the first fully connected layer, and $W^{P2}_n$, $b^{P2}_n$ are the weights and biases of the second fully connected layer. $I$ is the representation of the user's basic profile features drawn from the set {$B$}. 

The basic profile features used in the PEG module include demographic information such as age, gender, location, and educational background, as well as behavioral statistics such as the number of items presented, the number of items clicked, and the most interacted authors of the user, such as the most watched authors. All types of features are crucial for achieving optimal performance. More detailed discussions on the impact of these features are provided in the Experiment section. 

\subsubsection{Personalized Convolution}

After generating the convolution filters, we use these personalized filters to replace the original global filters in each TCN layer. An alternative approach is to use both the global and personalized filters and combine the results. However, we found that this method only provides marginal improvements compared to a complete replacement. Consequently, in the proposed CAIN, equation \ref{eq:conv} is rewritten as: 
\begin{equation} 
    cr_t = (ctxe_t \cup lhe_t) \times W^{PC} + b_c,
\label{eq:pconv}
\end{equation}

By using personalized convolution filters, the model can better capture the unique patterns of individual user behaviors, leading to more precise predictions.

\section{experiments}

To evaluate the performance of the proposed CAIN, we conducted extensive experiments using both a public dataset and an industrial dataset collected from the WeChat Channels platform. Additionally, we performed online A/B testing to assess the model's performance in a real-world environment. In this section, we provide a detailed description of our experimental setup, results, and analyses. 

\subsection{Experimental Setup}

\subsubsection{\textbf{Datasets}}

Our experiments were conducted using two datasets: one public dataset and one industrial dataset.

\textbf{Public Dataset}. We used the Taobao dataset, which is collected from the traffic logs of Taobao's recommendation system\footnote{https://tianchi.aliyun.com/dataset/649}. The dataset contains more than 100 million instances from over 1 million users within a period of 9 days. The average length of the user historical behavior sequence is 101, with the maximum length being 848. Since the dataset only includes instances with positive action feedback, we pre-processed the data similarly to previous methods \cite{pi2019practice, chen2022efficient} to generate positive and negative samples. The dataset was partitioned temporally, with the initial 8 days' data reserved for training and the data from the 9th day used as the test set. 

\textbf{Industrial Dataset}. This dataset was collected from user traffic logs on the WeChat Channels platform. It contains 13 billion instances from 0.3 billion users, collected over a period of 7 days. Instances with click feedback are treated as positive samples, while those without are considered negative samples. The average length of the user's historical behavior sequence is around 1500, with a maximum length of 2000. The dataset was partitioned temporally, with the initial 6 days' data reserved for training and the data from the 7th day used as the test set. 

\subsubsection{\textbf{Competitors}}

We compared the proposed CAIN to a series of state-of-the-art (SOTA) LSM methods. Our baseline model is built based on LAP \cite{hou2024cross}. Below, we detail the methods used in our comparisons:

\begin{itemize}

\item LAP \cite{hou2024cross}: Our initial baseline. It utilizes a three-level attention pyramid to refine the process of LSM.

\item SIM Soft \cite{pi2020search}: An early work that splits LSM into the GSU and ESU stages.

\item ETA \cite{chen2022efficient}: An approach that incorporates SimHash to achieve end-to-end learning of the GSU stage.

\item SDIM \cite{cao2022sampling}: A method that selects relevant items using the same hash signature across the GSU and ESU stages.

\item TWIN \cite{chang2023twin}: A method that adopts dimension compression to increase the consistency between GSU and ESU.

\end{itemize}

When implementing the above methods, we adopted the parameters and configurations reported in their original publications or open-source implementations.

\subsubsection{\textbf{Metrics}}

We conducted both offline and online experiments to evaluate the performance of the proposed CAIN model. For offline experiments, we used three widely recognized metrics in CTR prediction: Area Under the ROC Curve (AUC), Grouped Area Under the Curve (GAUC), and Logarithmic Loss (logloss). AUC and GAUC assess the model's pairwise ranking ability between positive and negative samples, while logloss measures the overall convergence of the training loss. In the online experiments, we evaluated the model's performance using industry-standard metrics: Click-Through Rate (CTR) and user stay time on presented items. Additionally, we considered online inference latency as a secondary metric to analyze the model's computational cost. 

\subsubsection{\textbf{Parameter Settings}} 

Our network architecture consists of two fully connected layers with dimensions set to 2048 and 1024, respectively. All input features are transformed into feature representations with an embedding size of 64 and concatenated before being fed into the first fully connected layer. We initialize the model parameters using the Xavier Initialization method \cite{glorot2010understanding} and optimize the model with the Adam optimizer \cite{kingma2014adam}, setting the learning rate to 0.001. The batch size is configured to 1536. All models are trained using the TensorFlow framework \cite{abadi2016tensorflow}. For the public dataset, we train the models on a single A100 GPU, while for the industrial dataset, we utilize 16 distributed A100 GPUs to handle the larger scale of data. 

\subsection{Model Analyses}
We designed experiments to analyze the effectiveness of the three key enhancements in CAIN: the TCN framework, the MSIA module, and the PEG module. All models in this section were trained using the industrial dataset, which is more suitable for LSM due to its substantial data volume and extended sequence length. 

\subsubsection{\textbf{Analyses of the TCN Framework}}

To understand the effectiveness of the TCN framework, we conducted an ablation study and analyzed its performance under different parameter settings. We also integrated the TCN framework with other LSM methods to evaluate its compatibility across various LSM backbones. The experiments in this section consider settings with only a single TCN layer, excluding the MSIA and PEG modules. 

\begin{figure}
    \centering
    \includegraphics[width=0.95\linewidth]{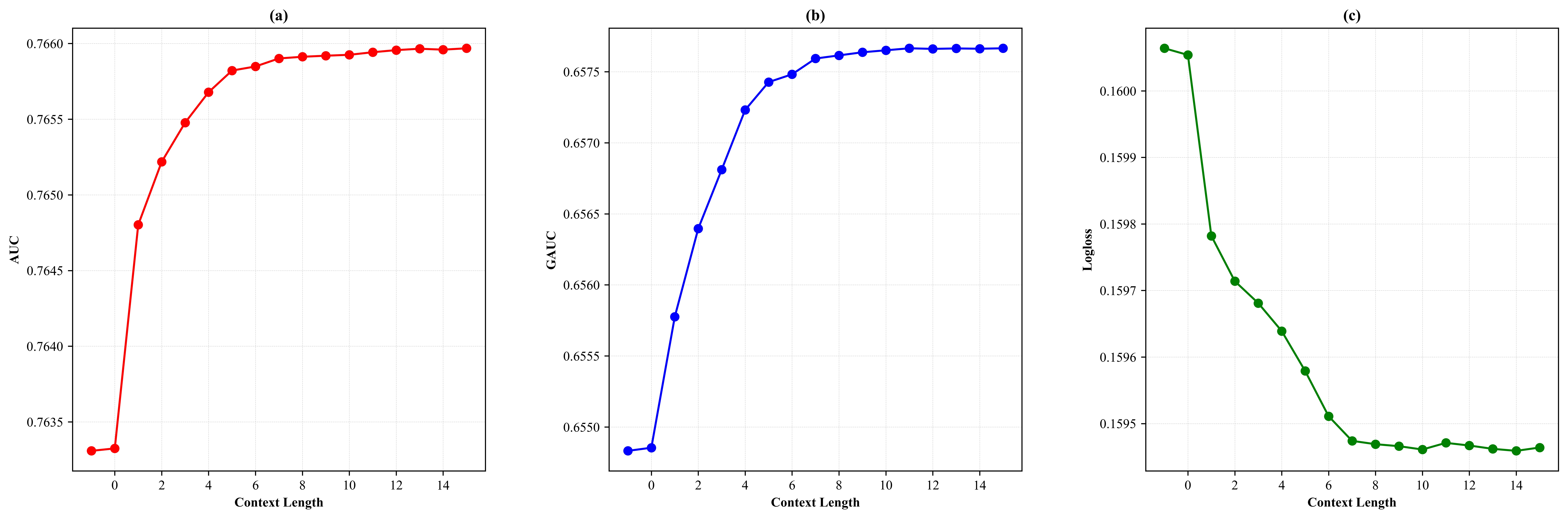}
    \caption{Performance comparison with different context lengths. }
    \label{fig:contextLength}
    \vspace{-0.5cm}
\end{figure}

\textbf{Influence of Context Length.} We conducted a set of experiments to analyze the performance of the model trained with different context lengths in the TCN framework. The results are shown in Figure \ref{fig:contextLength}. For experiments with a context length of -1, the TCN framework was not applied. For experiments with a context length of 0, no context was included, and the TCN layer degraded to a fully connected layer applied only to the center item. 

The results indicate that the TCN framework significantly improves the model's performance when the context length is greater than 0. The performance of the model with context lengths set to -1 and 0 showed little difference, suggesting that the improvements brought by the TCN framework do not stem from the additional computation introduced by the TCN layer. Instead, they primarily arise from the context information encoded in the context-aware representations output by the TCN layer, which helps better understand the user's intent regarding the items in the sequence.

The results also show that the model's performance continues to improve as the context length increases, peaking when the context length is set to a value between 7 and 15. This indicates that including more adjacent items in the context extraction provides a more comprehensive view of the changes in user interest around the center item. However, the performance starts to decrease when the context length continues to increase. This may be because the model becomes more difficult to converge, as behavior patterns within longer context lengths become more complex and unstable. Additionally, the information of the center item may be overwhelmed by its adjacent items, which could affect the subsequent attention module. For the remaining experiments, we set the context length of the first TCN layer to 7. 

\textbf{Comparison to Other Methods.} We conducted a set of experiments to compare the effectiveness of the TCN layer against the RNN layer \cite{hochreiter1997long,chung2014empirical} and Self-Attention \cite{vaswani2017attention}. In these experiments, we simply replaced the TCN layer with the compared methods. The results are presented in Table \ref{tab:ctxmethod}. 

\begin{table}[!htbp]
\center
\vspace{-0.3cm}
\caption{Performance comparison between different context extraction methods.}
\begin{tabular}{l|lll}
  \toprule
  Methods & AUC  & GAUC  & Logloss  \\ 
  \midrule
  RNN    & 0.76342  & 0.65494 & 0.15998 \\
  Self-Attention  & 0.76385     & 0.65516   & 0.15986 \\
  Ours & \textbf{0.76590} & \textbf{0.65759} & \textbf{0.15947} \\
  \bottomrule
\end{tabular}
\label{tab:ctxmethod}
\vspace{-0.3cm}
\end{table}

The results show that the proposed TCN framework outperforms models trained with RNN and Self-Attention. This is primarily because RNN and Self-Attention models are more difficult to converge, as they must not only learn the context of the center item but also determine the appropriate length of the context to include for a given item. In contrast, the context length can be explicitly controlled when using the TCN layer by adjusting the size of the convolution filters. This simplifies the modeling of context information and allows the model to focus more on learning the relevance between the output context-aware representations and the candidate items. It is also worth noting that RNN and Self-Attention methods are much more computationally costly than TCN and may not be suitable for deployment in models with lifelong sequence settings. 

\textbf{Backbone Substitution.} We conducted a set of experiments to incorporate the TCN framework with other LSM methods. In these experiments, the item representations used by these methods were replaced with the context-aware representations extracted by the TCN layer. The results are presented in Table \ref{tab:tcnLSMs}. 

\begin{table}[t]
    \centering
    \caption{Performance comparison of the TCN Framework incorporated with different LSM backbones.}
    \begin{tabular}{cccc}
    \toprule
     Methods & AUC & GAUC & Logloss \\
    \midrule
    \multicolumn{1}{l|}{SIM Soft} & 0.75022 & 0.64371 & 0.17268 \\
    \multicolumn{1}{l|}{SIM Soft$_{TCN}$} & \textbf{0.75209} & \textbf{0.64535} & \textbf{0.17181} \\
    \midrule
    \multicolumn{1}{l|}{ETA} & 0.74896 & 0.64238 & 0.17315 \\
    \multicolumn{1}{l|}{ETA$_{TCN}$} & \textbf{0.75108} & \textbf{0.64425} & \textbf{0.17165} \\
    \midrule
    \multicolumn{1}{l|}{SDIM} & 0.74657 & 0.64339 & 0.17304 \\
    \multicolumn{1}{l|}{SDIM$_{TCN}$} & \textbf{0.74851} & \textbf{0.64518} & \textbf{0.17194} \\
    \midrule
    \multicolumn{1}{l|}{TWIN} & 0.75211 & 0.64515 & 0.17208 \\
    \multicolumn{1}{l|}{TWIN$_{TCN}$} & \textbf{0.75409} & \textbf{0.64696} & \textbf{0.17089} \\
    \midrule
    \multicolumn{1}{l|}{LAP} & 0.76331 & 0.65483 & 0.16006 \\
    \multicolumn{1}{l|}{LAP$_{TCN}$ (Ours)} & \textbf{0.76590} & \textbf{0.65759} & \textbf{0.15947} \\
    \bottomrule
    \end{tabular}
    \label{tab:tcnLSMs}
    \vspace{-0.3cm}
\end{table}	

The results show that models trained with the proposed TCN framework consistently outperform their corresponding baselines. This improvement can be attributed to the context information embedded in the context-aware representations extracted by the TCN layer. This demonstrates the effectiveness and compatibility of the proposed TCN framework across different LSM methods. 

\subsubsection{\textbf{Analyses of the MSIA Module}}

We conducted a set of experiments to evaluate the effectiveness of the proposed MSIA module under various parameter settings, including the number of layers and the stride of the TCN layers. Additionally, we integrated the MSIA module with other LSM methods to assess its compatibility. 

\begin{figure}
    \centering
    \includegraphics[width=0.95\linewidth]{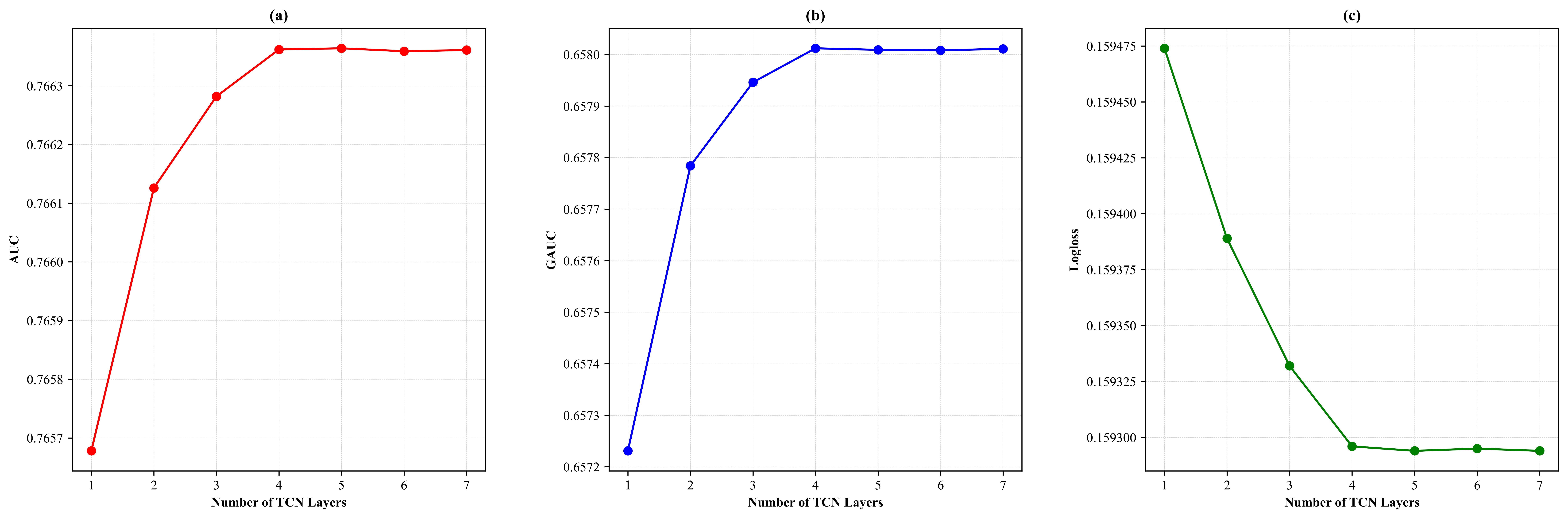}
    \caption{Performance comparison with different number of TCN layers. }
    \label{fig:MSIALayerDepth}
    \vspace{-0.5cm}
\end{figure}

\textbf{Influence of the Layer Depth.} We conducted a set of experiments to analyze the performance of the MSIA module with varying numbers of TCN layers. When the number of layers is set to 1, the MSIA module is not implemented. In these experiments, the context length and the stride of the convolution operation are set to match those of the first TCN layer for a fair comparison. The results are shown in Figure \ref{fig:MSIALayerDepth}.

The results indicate that the model's accuracy gradually increases as the number of TCN layers increases. This improvement is primarily due to the latter layers in the MSIA module introducing context-aware representations with a longer context length than the first layer. This helps the model capture user interest across a broader spectrum of context scopes. However, the model's performance peaks when the number of layers reaches 4 and does not improve further with additional layers. This may be because items too far from the center item contribute little to understanding the user's intent regarding the center item. Consequently, the information gain from increasing the context length becomes limited. 

It is important to note that increasing the context length by adding more TCN layers is very different from increasing the size of the convolution filters in the first TCN layer (tested in Figure \ref{fig:contextLength}). With only one TCN layer of large filter size, the model relies solely on this single representation with a long context length, making it harder to converge when the length is very long. In contrast, by stacking more TCN layers and their corresponding attention modules and combining their outputs, the model can obtain information from different context scopes, including those with shorter context lengths in the initial TCN layers. This allows the model to learn both immediate behavior changes that are more relevant to the center item and interest changes that reveal the influence of the center item over a longer period. At the same time, the representativeness of the output from the latter layers increases due to the non-linearity of the TCN operations, which also contributes to the final performance. We set the number of layers to 4 in the remaining experiments. 

\begin{figure}
    \centering
    \includegraphics[width=0.95\linewidth]{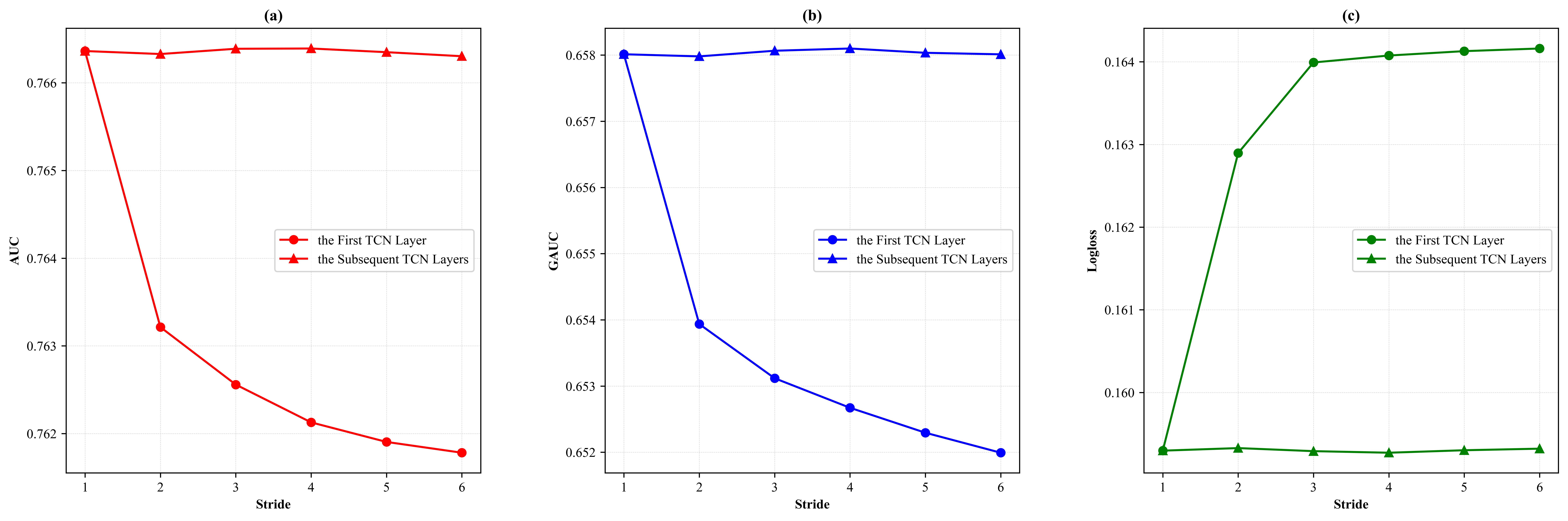}
    \caption{Performance comparison with different convolution strides. }
    \label{fig:MSIAStride}
    \vspace{-0.7cm}
\end{figure}

\textbf{Influence of the Stride Setting.} We conducted a set of experiments to analyze the model's performance when the stride of the first TCN layer or the stride of the subsequent TCN layers is set to a value greater than 1. The results are shown in Figure \ref{fig:MSIAStride}.

The results indicate that the model's performance deteriorates as the stride of the first TCN layer increases. The performance is even worse than our baseline model without the TCN framework and the MSIA module. The side effect of increasing the stride of the convolution operation is that different items in the input sequence share representations in the output sequence. This implies that some unique information of the individual center items in the input may be lost. When performing attention in a lifelong sequence, this information may be crucial to ensure that the most relevant items are retained until the final level of the attention pyramid (or the ESU stage for other LSM methods). 

In contrast, the results show little performance difference among models trained with a convolution stride of 1 to 6 in the subsequent TCN layers. This suggests that the information of the center item is less important for the latter TCN layers, as it is already included in the output of the first TCN layer and the major attention. The subsequent TCN layers and auxiliary attentions are more focused on providing representations with longer context lengths. It is worth noting that although there is no accuracy gain from increasing the convolution stride, larger strides can significantly reduce the length of the TCN layer outputs, thereby greatly reducing the computational cost in the auxiliary attentions and subsequent layers. Therefore, we set the stride of the subsequent TCN layers to 4. 

\textbf{Backbone Substitution.} We conducted a set of experiments to test the compatibility of the MSIA module with other LSM methods. In these experiments, the item representations used by these methods were replaced with the context-aware representations extracted by the first TCN layer in the MSIA module. The interest representations from the subsequent MSIA layers were simply concatenated with those from the first layer. The results are presented in Table \ref{tab:MSIA_LSMs}.
		
\begin{table}[t]
    \centering
    \caption{Performance comparison of the MSIA module incorporated with different LSM backbones.}
    \begin{tabular}{cccc}
    \toprule
     Methods & AUC & GAUC & Logloss \\
    \midrule
    \multicolumn{1}{l|}{SIM Soft} & 0.75022 & 0.64371 & 0.17268 \\
    \multicolumn{1}{l|}{SIM Soft$_{MSIA}$} & \textbf{0.75287} & \textbf{0.64643} & \textbf{0.17138} \\
    \midrule
    \multicolumn{1}{l|}{ETA} & 0.74896 & 0.64238 & 0.17315 \\
    \multicolumn{1}{l|}{ETA$_{MSIA}$} & \textbf{0.75152} & \textbf{0.64491} & \textbf{0.17119} \\
    \midrule
    \multicolumn{1}{l|}{SDIM} & 0.74657 & 0.64339 & 0.17304 \\
    \multicolumn{1}{l|}{SDIM$_{MSIA}$} & \textbf{0.74906} & \textbf{0.64603} & \textbf{0.17128} \\
    \midrule
    \multicolumn{1}{l|}{TWIN} & 0.75211	& 0.64515 & 0.17208 \\
    \multicolumn{1}{l|}{TWIN$_{MSIA}$} &  \textbf{0.75478} & \textbf{0.64796} & \textbf{0.17033} \\
    \midrule
    \multicolumn{1}{l|}{LAP} & 0.76331 & 0.65483 & 0.16006 \\
    \multicolumn{1}{l|}{LAP$_{MSIA}$ (Ours)} &  \textbf{0.76639} & \textbf{0.65810} & \textbf{0.15927} \\
    \bottomrule
    \end{tabular}
\label{tab:MSIA_LSMs}
\vspace{-0.4cm}
\end{table}		

The results show that the performance of the models trained with the MSIA module is better than those trained without it. This conclusion is consistent across different LSM methods. Additionally, the performance of the models trained with the MSIA module is even better than their corresponding performance in Table \ref{tab:tcnLSMs}. This demonstrates the effectiveness of the MSIA module and its robustness to changes in the LSM backbones. 

\subsubsection{\textbf{Analyses of the PEG Module}} 

We conducted an ablation study to assess the effectiveness of the PEG module. Additionally, we analyzed the impact of input selection and the method of aggregating outputs on the final performance. 

\textbf{Ablation Study.} We conducted a set of experiments to analyze the effectiveness of the PEG module when implemented on the first TCN layer, the subsequent TCN layers in the MSIA module, or both. The results are presented in Table \ref{tab:PEG_layers}. In this experiment, we used a comprehensive set of input features for the PEG module. When the PEG module is implemented, the original global convolution filters are replaced by those generated by the PEG module. 

\begin{table}[t]
    \centering
    \caption{Performance comparison with PEG module implemented on different TCN layers. }
    \begin{tabular}{cccc}
    \toprule
     TCN Layers used PEG & AUC & GAUC & Logloss \\
    \midrule
    \multicolumn{1}{l|}{MSIA} & 0.76639 & 0.65810 & 0.15927 \\
    \midrule
    \multicolumn{1}{l|}{MSIA + PEG for 1st layer} & 0.76704 & 0.65919 & 0.15891 \\
    \midrule
    \multicolumn{1}{l|}{MSIA + PEG for latter layer} & 0.76676 & 0.65868 & 0.15901 \\
    \midrule
    \multicolumn{1}{l|}{MSIA + PEG} & \textbf{0.76718} & \textbf{0.65965} & \textbf{0.15883} \\
    \bottomrule
    \end{tabular}
\label{tab:PEG_layers}
\vspace{-0.5cm}
\end{table}

The results show that the PEG module further improves the model's performance, with the improvement being more significant when the PEG module is implemented for all TCN layers in the MSIA module. Additionally, the improvement in the GAUC metric is greater than the improvement in the AUC metric. This indicates that the PEG module has enhanced the model's ability to produce more personalized results, thereby significantly improving the ranking results for individual users.

\textbf{Influence of the Input Features.} We conducted a set of experiments to analyze the impact of the three types of input features used in the PEG module. The results are shown in Table \ref{tab:PEG_inputFea}.

\begin{table}[t]
    \centering
    \caption{Performance comparison with different input features for the PEG module. }
    \begin{tabular}{cccc}
    \toprule
     Input features & AUC & GAUC & Logloss \\
    \midrule
    \multicolumn{1}{l|}{Demographic information} & 0.76664 & 0.65856 & 0.15905 \\
    \midrule
    \multicolumn{1}{l|}{Behavioral statistics} & 0.76669 & 0.65889 & 0.15899 \\
    \midrule
    \multicolumn{1}{l|}{Most interacted authors} & 0.76681 & 0.65893 & 0.15892 \\
    \midrule
    \multicolumn{1}{l|}{All features} & \textbf{0.76718} & \textbf{0.65965} & \textbf{0.15883} \\
    \bottomrule
    \end{tabular}
\label{tab:PEG_inputFea}
\vspace{-0.3cm}
\end{table}

The results indicate that while the best performance is achieved when all features are included, the feature of most interacted authors contributes the most to the final performance. This is expected, as these features are highly unique to different users, making them the most personalized in describing user interests. However, even when the most interacted authors are included, adding demographic and statistical features still improves the model's accuracy. This may be because, for less active users who do not have many positive interactions with authors, these additional features help the model generalize among users with similar basic profiles. This demonstrates the importance of including both personalized and generalized features as input to the PEG module to generate convolution filters that cater to all types of users. 

\textbf{Influence of the Aggregation Method.} We conducted a set of experiments to compare the effectiveness of the PEG module under different aggregation methods. The original implementation replaces the results of the global filters with those from the personalized filters generated by the PEG module. In addition to this, we tested the performance when results from both filters are used, and when results from different filters are summed or concatenated to produce the final output. The results are shown in Table \ref{tab:PEG_agg}.

\begin{table}[t]
    \centering
    \caption{Performance comparison with different aggregation methods applied in the PEG module. }
    \begin{adjustbox}{max width=1.0\linewidth}
    \begin{tabular}{cccc}
    \toprule
     Methods & AUC & GAUC & Logloss \\
    \midrule
    \multicolumn{1}{l|}{Replace} & 0.76718 & 0.65965 & 0.15883 \\
    \midrule
    \multicolumn{1}{l|}{Sum} & 0.76716 & 0.65962 & 0.15889 \\
    \midrule
    \multicolumn{1}{l|}{Concat} & 0.76723 & 0.65967 & 0.15880 \\
    \bottomrule
    \end{tabular}
    \end{adjustbox} 
\label{tab:PEG_agg}
\vspace{-0.5cm}
\end{table}

The results indicate that there is no significant difference between the various methods of aggregating the outputs. Using only the outputs from the personalized filters is sufficient to provide robust performance. This may be because the information in the representations extracted by the global filters is already contained in those extracted by the personalized filters, as the inputs to the PEG module also include more generalized features. Therefore, the information gain from including the original representations in the final output is limited and does not further improve the model. 

\subsection{Overall Performance}

The final performance of the proposed CAIN, along with its comparison to previous LSM methods, is presented in Table \ref{table:CAIN_overall}. Note that when the MSIA and PEG modules are not implemented, CAIN consists solely of the TCN framework with a single TCN layer.

\begin{table}[t]
    \centering
    \caption{Overall performance comparison.}
    \begin{adjustbox}{max width=1.0\linewidth}
    \begin{tabular}{ccccccc}
    \toprule
        \multirow{2}*{Methods} & \multicolumn{3}{c}{Industrial.} & \multicolumn{3}{c}{Public.} \\
        \cmidrule(lr){2-4}\cmidrule(lr){5-7}
        & AUC & GAUC & Logloss  & AUC & GAUC & Logloss \\
        \midrule
        SIM Soft    & 0.75022 & 0.64371 & 0.17268 & 0.87872 & 0.78671 & 0.14919 \\
        ETA         & 0.74896 & 0.64238 & 0.17315 & 0.87798 & 0.78573 & 0.14932 \\ 
        SDIM        & 0.74657 & 0.64339 & 0.17304 & 0.87705 & 0.78509 & 0.14943  \\
        TWIN        & 0.75211 & 0.64515 & 0.17208 & 0.87971 & 0.78752 & 0.14867 \\
        LAP         & 0.76331 & 0.65483 & 0.16006 & 0.88513 & 0.79423 & 0.14809 \\
        \cmidrule(lr){1-7}		
        CAIN $w/o$ MSIA and PEG & 0.76590 & 0.65759 & 0.15947 & 0.88664 & 0.79589 & 0.14761  \\
        CAIN $w/o$ PEG           & 0.76639 & 0.65810 & 0.15927 & 0.88696 & 0.79621 & 0.14743 \\
        CAIN          & \textbf{0.76718} & \textbf{0.65965} & \textbf{0.15883} & \textbf{0.88748} & \textbf{0.79714} & \textbf{0.14696} \\
        \bottomrule
    \end{tabular}
    \end{adjustbox}
    \vspace{-0.5cm}
    \label{table:CAIN_overall}
\end{table}		

The results demonstrate that across all metrics and datasets, the proposed CAIN consistently outperforms the other methods. Notably, the margin of improvement on the industrial dataset is greater than that on the public datasets. This is primarily because the sequence length and data volume are smaller for the public datasets. Therefore, the performance on the industrial dataset provides a more accurate illustration of the model's capability to handle complex real-world environments.

\subsection{Online Performance}

To further validate the efficiency of the proposed CAIN, we conducted an online A/B test to assess its performance in real-world industrial scenarios. We collected user feedback from the online platform for metrics calculation over a period of seven days. Compared to the control group, where the baseline LAP is implemented, the proposed CAIN in the experimental group received significantly better user feedback, with a 1.93\% increase in stay time and a 3.43\% increase in CTR. Moreover, the inference latency of the proposed CAIN is only 8 ms longer than the baseline model, owing to the light and fast nature of the TCN operations. This slight increase in latency is negligible compared to the substantial improvement in recommendation quality. 

\section{conclusions}

In this paper, we propose the Context-Aware Interest Network (CAIN) to incorporate context information from adjacent items in lifelong sequential modeling (LSM). The proposed CAIN is the first model in the field to integrate the Temporal Convolutional Network (TCN) \cite{bai2018empirical}, achieving efficient and fast context extraction throughout the lifelong sequence. Building on this framework, we introduce the Multi-Scope Interest Aggregator (MSIA) module, which contains multiple stacked TCN layers and their corresponding attention modules to extract and combine context-aware interest representations across various context scopes. Additionally, we propose the Personalized Extractor Generation (PEG) module to generate convolution filters for individual users, replacing the global filters in the TCN layers to achieve more user-specific responses. 

We conducted experiments on both public and industrial datasets. The results demonstrate that CAIN outperforms previous LSM methods on both datasets. Furthermore, we conducted online A/B testing, where CAIN significantly outperformed the initial baseline. CAIN now serves as the new baseline on our platform. For future work, we aim to explore ways to combine more context extractors and include context information to enhance the representation of candidate items as well.

\balance
\bibliographystyle{ACM-Reference-Format}
\bibliography{reference}


\end{document}